\documentclass[a4paper]{article}

\usepackage{INTERSPEECH2018}

\usepackage[]{siunitx}

\usepackage{cite}
\usepackage{graphicx}
\DeclareGraphicsExtensions{.png}
\usepackage{multirow}

\usepackage{times}
\usepackage{psfrag}
\usepackage{color}
\usepackage{colortbl}
\usepackage[varg]{txfonts}
\usepackage{mathptmx}	

\def\min{\mathop{\rm min}}

\def\tr{\mathop{\rm tr}}

\title{Recognizing Overlapped Speech in Meetings:\\A Multichannel Separation Approach Using Neural Networks}
\name{Takuya Yoshioka,  Hakan Erdogan, Zhuo Chen, Xiong Xiao, and Fil Alleva}
\address{
  Microsoft AI and Research, One Microsoft Way, Redmond, WA, USA}
\email{\{tayoshio, hakan.erdogan, zhuc, xioxiao, fil\}@microsoft.com}

\begin{document}

\maketitle
\begin{abstract}
The goal of this work is to develop a meeting transcription system that can recognize speech even when 
utterances of different speakers are overlapped. 
While speech overlaps have been regarded as a major obstacle in accurately transcribing meetings, 
a traditional beamformer with a single output has been exclusively used 
because previously proposed speech separation techniques have critical constraints for application to real meetings. 
This paper proposes a new signal processing module, called an unmixing transducer, and describes its implementation using a windowed BLSTM. 
The unmixing transducer has a fixed number, say $J$, of output channels, where $J$ may be different from the number of meeting attendees, and 
transforms an input multi-channel acoustic signal into $J$  time-synchronous audio streams. 
Each utterance in the meeting  is separated and emitted from one of the output channels. 
Then, each output signal can be simply fed to a speech recognition back-end for segmentation and transcription. 
Our meeting transcription system using the unmixing transducer outperforms a system based on a state-of-the-art neural mask-based beamformer by 10.8\%. 
Significant improvements are observed in overlapped segments. 
To the best of our knowledge, 
this is the first report that applies overlapped speech recognition to unconstrained real meeting audio.
\end{abstract}

\section{Introduction}
Automatic speech recognition (ASR) technology has made a significant stride over the past decade, achieving human parity in some domains~\cite{Xiong16,Amodei15}. 
However, when it comes to dealing with speech overlaps, the machines still lag far behind humans. 
Our brains can attend to one speaker in a noisy multi-talker environment and recognize what he/she has spoken even when 
his/her voice is overlapped by utterances of other speakers, as demonstrated by 
the cocktail party effect. 
By contrast, 
the current ASR systems fail miserably when utterances of two or more speakers overlap. 
Computational implementation of the ability of transcribing individual utterances that may or may not be overlapping in the multi-talker settings 
will be a cornerstone of a range of far-field conversation transcription systems, e.g., for meetings~\cite{Fiscus05,Hain12,Hori12,Renals17} and doctor-patient dialogs~\cite{Edwards17,Chiu17}.

In this paper, 
we develop a multi-microphone meeting transcription system that can recognize overlapped speech.
While speech recognition in the meeting space has a long history of research, 
most systems developed in the past was not able to handle speech overlaps. 
Overlap segments account for 10+\% of the speaking time~\cite{Cetin06}, which is too much to ignore.

Challenges that need to be overcome for the ASR systems to be able to recognize overlapped speech in practical far-field settings include 
an unknown and varying number of speakers, unknown speaker identities, 
unknown speech activity segments, the presence of background noise and reverberation, and online operation. 
Numerous approaches have been proposed for speech separation or overlapped speech recognition, 
such as
%computational auditory scene analysis, 
independent component analysis, 
time-frequency bin clustering, 
and deep neural networks~\cite{Hershey16,Drude17, Chen17a,Zhang17,Zmolikova17}.
However, previous research in these areas was mostly conducted in \textit{in vitro} settings, e.g., 
by supposing prior knowledge of the number of meeting attendees.  
%by assuming acoustic environments with little background noise or reverberation. 
While techniques like overlap detection or speaker counting may help close the gap between the laboratory and practical settings, 
orchestrating many different error-prone components is not so easy as it appears. 

\begin{figure}
\centering
\includegraphics[scale=0.4]{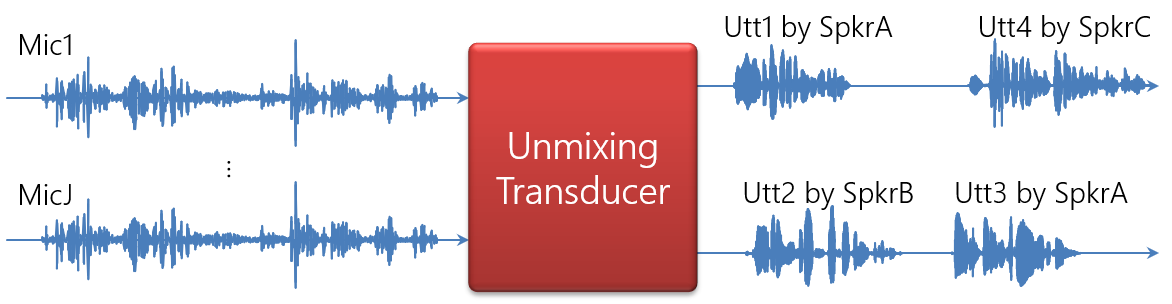}
\vspace{-2em}
\caption{Unmixing transducer converts an input $J$-channel signal into a fixed number of time-synchronous audio streams. Each utterance ``spurts'' from one of 
the output channels. The ouptut from each channel is fed to an ASR back-end for segmentation and recognition.}
\label{fig: ut}
\vspace{-2em}
\end{figure}

We show that those challenges can be addressed by 
a novel signal processing module, called an \textit{unmixing transducer}, followed by an array of ASR back-ends.
The unmixing transducer continuously receives microphone signals and generates a fixed number of time-synchronous audio streams as illustrated 
in Fig.~\ref{fig: ut}. 
The acoustic signal of each utterance found in the input ``spurts'' from one of the output channels. 
When the number of active speakers is fewer than that of the outputs, 
the extra channels generate zero-valued signals. 
The signal from each output channel is segmented and transcribed by the back-end recognizer connected to that channel.
%For applications where latency is not critical, 
%a second decoding pass may be run with bidirectional models 
%on the speech segments found by the first pass. 

The unmixing transducer is implemented by extending our recently proposed method~\cite{Yoshioka18}, which is based on acoustic beamformers driven by a multi-microphone speech separation neural network using permutation invariant training (PIT)~\cite{Kolbaek17}. 
Our extensions include the use of a windowed BLSTM for handling long audio streams, 
a new model architecture suitable for beamforming, 
improved feature normalization taking account of phase wrapping, 
and addition of spherically isotropic random noise to training data. 
Dereverberation is also performed by using the weighted prediction error (WPE) method~\cite{Yoshioka12c,Li17} to further improve the reverberation robustness.

%The effectiveness of the proposed  meeting transcription system is evaluated by using seven-microphone recordings of inhouse meetings.
Our proposed meeting transcription system is shown to work reasonably well for real meeting data 
that we collected at our speech group meetings. 
Compared with a state-of-the-art neural mask-based beamformer, the proposed unmixing transducer is demonstrated to be particularly effective in dealing with overlaps.

\section{Unmixing Transducer}

This section describes what the functionality of the unmixing transducer is and how it is fulfilled in our proposed system. 
For simplicity and conciseness, we assume the maximum number of overlaps to be two at each time instant, which is true 98+\% of the time according 
to \cite{Cetin06}. Extension to more overlaps is straightforward. 
No assumption is made on the total number of meeting attendees.

%As a prerequisite, 
%we define what we mean by a term utterance (in a rather ambiguous way). In this paper, an utterance refers to a person's speech activity that is localized in time. 
%Metaphorically speaking, it is a spurt discharged from a geyser, where the geyser compares to a speaker participating in the conversation to be transcribed. 
%The meeting participants are assumed to get active in a way that no more than two speakers are active at the same time. 

\subsection{Problem}
\vspace{-.4em}

As shown in Fig.~\ref{fig: ut}, 
the unmixing transducer receives acoustic signals from a microphone array, in which utterances from different speakers are reverberated and mixed.
It separates each utterance from coincident utterances, if any, and emits the separated signal from either of its two output channels. 
Each utterance should not be broken up and distributed to multiple output channels. 
For time segments where zero or one speaker is active, the extra output channel yields a zero-valued signal. 
In this way, it always produces two time-synchronous audio streams.

While the above description may be sufficient, 
we provide a more formal definition of the problem in the following. 
We represent signals in the time-frequency domain by denoting time and frequency by $t$ and $f$, respectively.
For each utterance in the meeting, we consider a padded utterance signal, $u_{k, tf}$, where $k$ is the utterance index. 
Each padded utterance signal is as long as the meeting and is created by taking 
the time-localized utterance signal as measured by a reference (e.g., the first) microphone and padding the inactive time segments with zero. 
Now, we consider mapping $\varphi : \{0, \cdots, K-1\} \mapsto \{0, 1\} $ with $K$ being the total number of the utterances. 
This defines which output channel for each utterance to go. 
The inverse of the mapping can also be defined as $\varphi^{-1}[i] =  \{k;   \varphi[k] = i\}$, 
which, for output channel $i$, returns the set of the utterances that are mapped to $i$ by $\varphi$.
We call $\varphi$ a nonmixing mapping when it meets the following condition for all $t$ values\footnote{Condition $u_{k, tf} \neq 0, ~\exists f$ is assumed to
be equivalent to the $k$th utterance being active at frame $t$. The utterance is regarded as inactive iff $u_{k, tf} = 0, ~\forall f$.}: 
\begin{align}
u_{k, tf} \neq 0, ~\exists f \implies u_{k', tf} = 0, ~\forall f,~\forall k' \in \varphi^{-1}[ \varphi[k] ] \setminus \{k\}.
\label{eq: valid_mapping}
\end{align}
Nonmixing mappings keep each utterance isolated from each other, ensuring that 
the following superimposed signal consists of at most one utterance at any time: 
\begin{align}
%s_{i, tf} = \sum_{\substack{0 \leq k < U \\ \varphi[k]=i}} u_{k, tf}. 
s_{i, tf} = \sum_{0 \leq k < K , ~\varphi[k]=i} u_{k, tf},~~~ i \in \{0, 1\}. 
\end{align}
We want to find such ``unmixed'' signals for some nonmixing mapping.

\subsection{Masking approach}
\vspace{-.4em}
We start by a simple approach using spectral masking. 
While the system we eventually develop does not  perform masking, the mask estimation processing constitutes an essential element of our system. 
Let $x_{j, tf}$ and $y_{i, tf}$ denote the $j$th input to and $i$th output from the unmixing transducer, respectivey. 
For each output channel $i$, spectral mask $m_{i, tf}$, whose value is bounded between 0 and 1, is estimated. 
The mask is applied to the reference microphone, with index R, to obtain 
the $i$th output signal as $y_{i, tf} = m_{i, tf} x_{\text{R}, tf}$. 
We want $y_{i, tf}$ to be close to $s_{i, tf}$ which can be derived with some nonmixing mapping.

\begin{figure}
\centering
\includegraphics[scale=0.48]{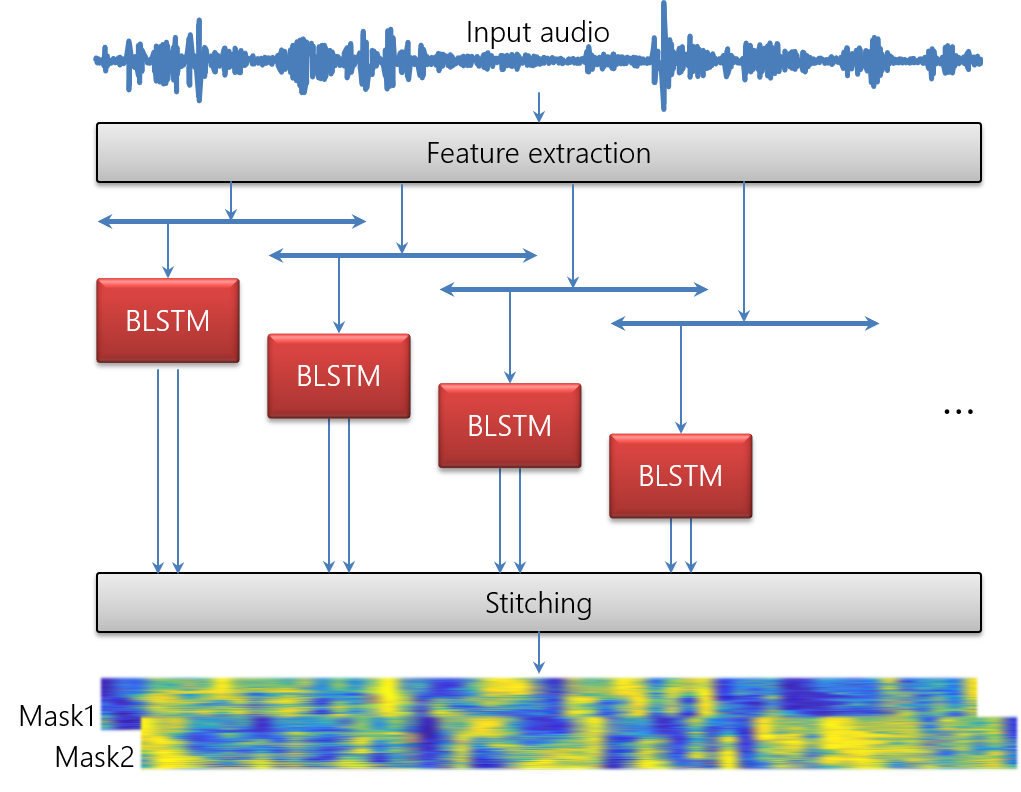}
\vspace{-2em}
\caption{Mask estimation with windowed BLSTM.}
\label{fig: blstm}
\vspace{-1em}
\end{figure}

We propose to calculate the spectral masks with a windowed BLSTM\footnote{The windowed BLSTM was previously proposed for acoustic modeling~\cite{Mohamed15}.} as illustrated in Fig. \ref{fig: blstm}. 
The incoming audio signals, which may last for tens of minutes to hours in typical office meetings, 
are broken up into overlapping time windows.  
Our system uses a 2.4-sec sliding window with a 75\% overlap. 
Feature vectors in each window are fed to a speech separation BLSTM that yields the spectral masks for the respective window.
The spectral masks from the adjacent windows are ``stitched'' to form sequences of spectral masks in a way 
that does not split an utterance to different output channels (see Sec. \ref{subsubsec: stitch} for details).

The windowed BLSTM is chosen for two reasons. First, 
the model can be trained on a collection of short (i.e., not as long as a typical meeting) speech mixtures, which can be easily created by simulation. 
Secondly, it can efficiently capture temporal feature dependency, which 
is critical for the separated speaker signals not to be swapped within each window.

\subsubsection{Input features}
\vspace{-.2em}

As input to the BLSTM, we make use of both spectral and spatial features. 
The magnitude spectrum of the reference microphone is used as the spectral features. 
As regards the spatial features, 
inter-microphone phase differences (IPDs) relative to the reference microphone are used. 
%Thus, the features for time frame $t$ are written as 
%$(a_{tf}, p_{2, tf}, \cdots, p_{J, tf})_{0 \leq f < F}$,
%where $a_{tf}$ and $p_{j, tf}$ are the spectral and spatial features, respectively, computed as follows:
%\begin{align}
%a_{tf} =& |x_{\text{R}, tf}| \\
%p_{j, tf} =& \angle( x_{j, tf} ) - \angle(x_{\text{R}, tf}), ~~~j \neq \text{R}.
%\end{align}
%Here, $F$ denotes the number of frequency bins, and $J$ is that of the microphones. 
All the features are mean-normalized by using a rolling window of four seconds. 
Unlike in \cite{Yoshioka18}, 
to prevent aliasing at the $\pi$/-$\pi$ boundary, 
the argument operation is performed 
after the mean normalization processing. Thus, the IPD features are calculated as 
\begin{align}
\text{Arg} \Biggl( \frac{x_{j, tf}}{x_{\text{R}, tf}} - E_{\tau}  \Bigl( \frac{x_{j, \tau f}}{x_{\text{R}, \tau f}} \Bigr) \Biggr), ~~~j \neq \text{R}, 
\end{align}
where the time averaging operator, $E_{\tau}$, is applied over the normalization window.

%we normalize the IPD features on the unit circle as follows:
%\begin{align}
%p_{j, tf} \gets \angle \Bigl(   e^{ \text{j} p_{j, tf}} - \sum_{t} e^{ \text{j} p_{j, tf} } \Bigr), 
%\end{align}
%where $p_{j, tf}$ denotes the IPD between the $j$th microphone and the reference microphone while 
%$\text{j}$ stands for the imaginary unit. 

\subsubsection{Training}
\vspace{-.2em}

The BLSTM is trained with PIT so that the resultant model can consistently 
assign each separated utterance to either channel within a window. 
Our training set comprises simulated multi-channel signals of up to 10 seconds. 
Each signal can be a single utterance or a mixture of two utterances with different 
lengths, levels, and reverberations, corrupted by background noise. 
The PIT loss for the $l$th training sample is defined as 
\begin{align}
\min_{( j_0, j_1 ) \in  \{ (0, 1), (1, 0) \}} \sum_{i=0}^{1} \sum_{tf}  \Bigl(  m_{i, tf}^{(l)} \bigl| x_{\text{R}, tf}^{(l)} \bigr|  - \bigl| s_{j_i, tf}^{(l)} \bigr|  \Bigr)^2, 
\label{eq: pit}
\end{align}
where $m_{i, tf}^{(l)}$, $x_{\text{R}, tf}^{(l)}$, and $s_{i', tf}^{(l)}$ are 
the $i$th output from the model, the reference microphone signal, and the 
$i'$th source signal as measured at the reference microphone position, respectively. 
For the training samples involving only one utterance, $s_{i', tf}^{(l)} = 0$  for the extra source. 
Further details are described in Section \ref{subsec: train}.

\subsubsection{Stitching adjacent windows}
\label{subsubsec: stitch}
\vspace{-.2em}

Because the PIT-trained network has no specific preference as to the ordering of the separated signals, 
the permutations of the separated signals need to be aligned across the windows at test time. 
Suppose that the permutations have already been determined up to the previous window. 
To decide the output signal permutation for the current window, we calculate the cost of each possible permutation and pick the one that provides the lower cost. 
The cost is defined as the sum of the squared differences between the separated signals of the adjacent windows, where the sum is computed 
over the  overlapping frames. 
%\begin{align}
%e_{0}^{(w+1)} = \sum_{t, f} \Bigl|u_{0, tf}^{(w)} - u_{0, tf}^{(w+1)} \Bigr|^2 + \Bigl|u_{1, tf}^{(w)} - u_{1, tf}^{(w+1)} \Bigr|^2 \\
%e_{1}^{(w+1)} = \sum_{t, f} \Bigl|u_{0, tf}^{(w)} - u_{1, tf}^{(w+1)} \Bigr|^2 + \Bigl|u_{1, tf}^{(w)} - u_{0, tf}^{(w+1)} \Bigr|^2
%\end{align}

After the permutation alignment processing, the masks for the nonoverlapping frames of the current window are used. 
This minimizes the processing latency while being not optimal in terms of accuracy.

\subsection{Beamforming approach}
\vspace{-.4em}

While spectral masking provides perceptually enhanced sounds, 
there is a shared belief that the processing artifacts created by masking are detrimental to the current ASR technology. 
To overcome this drawback, a mask-based beamforming approach was proposed~\cite{Yoshioka15b,Heymann15} and showed the state-of-the-art results in
far-field ASR tasks~\cite{Heymann17,Boeddeker18}. 

With beamforming, 
the output signals are computed as 
\begin{align}
y_{i, tf} = \bm{w}_{c, i, f}^H \bm{x}_{tf}, 
\end{align}
where $\bm{w}_{c, i, f}$ is a beamformer coefficient vector for output channel $i$, 
$\bm{x}_{tf}$ is a vector stacking the microphone signals, 
and $c$ is the window index which $y_{i, tf}$ belongs to. 
By using the MVDR method~\cite{Souden10,Erdogan16}, 
the  optimal beamfomer is obtained as 
$\bm{w}_{c, i, f} = \bm{\Psi}_{c, i, f}^{-1} \bm{\Phi}_{c, i, f} \bm{e} / \varrho_{c, i, f}$, 
where the normalization term, $\varrho_{c, i, f}$, is calculated as $\varrho_{c, i, f} =  \tr  (\bm{\Psi}_{c, i, f}^{-1} \bm{\Phi}_{c, i, f} )$. 
Here, $\bm{e}$ is the $J$-dimensional standard basis vector with 1 at the reference microphone position. 
The two matrices, $\bm{\Phi}_{c, i, f}$ and $\bm{\Psi}_{c, i, f}$, represent
the spatial covariance matrix of the utterance to be output from the $i$th channel (which may be referred to as the target utterance)
and that of the sounds overlapping the target.  
These matrices were previously estimated as weighted spatial covariance matrices of the microphone signals, 
where each microphone signal vector was weighted by $m_{i, tf}$ for the target or  $1-m_{i, tf}$ for the interference~\cite{Yoshioka18}. 
In the following, we propose an improved spatial covariance matrix estimator using a different model architecture.

\subsubsection{Speech-speech-noise architecture}
\vspace{-.2em}

The spatial covariance matrix estimator that uses $1-m_{i,tf}$ as the interference mask is not very accurate. 
This is because the trained speech separation network cares only about the target signal estimation accuracy, as evident from \eqref{eq: pit}.

A more accurate estimate of the interference spatial covariance matrix can be obtained 
by explicitly factorizing it to 
the spatial covariance matrix of the other talker's speech and that of the background noise, $\bm{\Phi}_{c, \text{N}, f}$,  as follows:
\begin{align}
\bm{\Psi}_{c, i, f} = \bm{\Phi}_{c, \Bar{i}, f} + \bm{\Phi}_{c, \text{N}, f}, 
\end{align}
where $\Bar{i} = 0$ for $i = 1$ and $\Bar{i} = 1$ for $i = 0$. 

\begin{figure}
\centering
\includegraphics[scale=0.48]{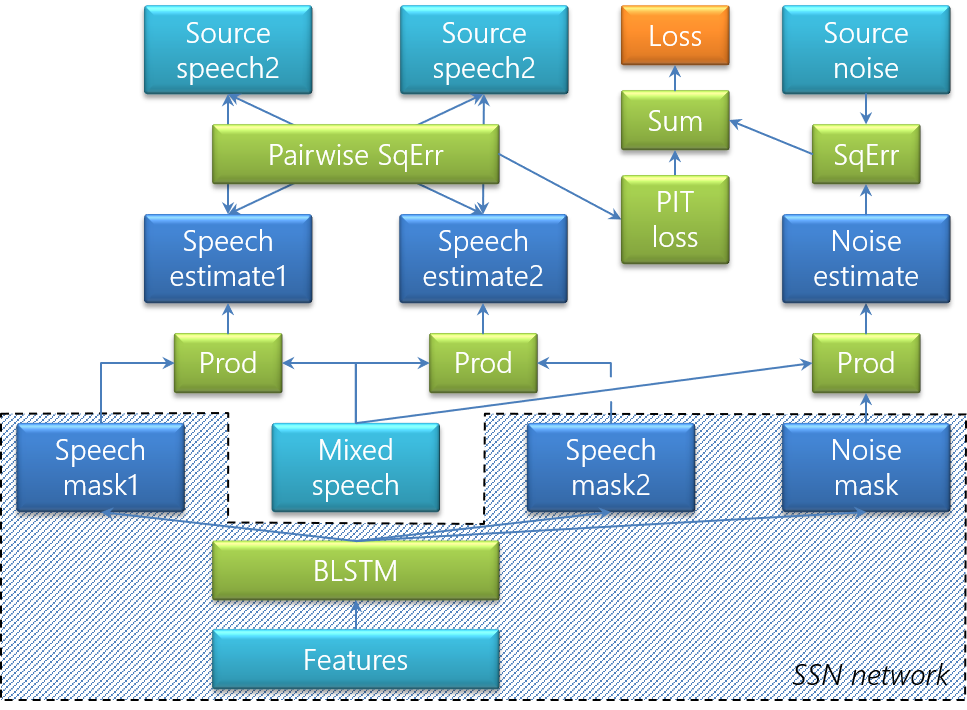}
\vspace{-.7em}
\caption{SSN model and the network for training it.}
\label{fig: SSN}
\vspace{-1em}
\end{figure}

To obtain $\bm{\Phi}_{c, \text{N}, f}$, we add another output channel to the separation network so that the noise masks can also be obtained. 
Figure \ref{fig: SSN} shows a diagram of the new model, 
called the speech-speech-noise (SSN) model\footnote{SSN also stands for Speech Separation Network.}, and the computational network for training it. 
As shown in the diagram, we apply the PIT framework only to the first two (i.e., speech) output channels. 
The loss function is defined as the sum of the PIT loss and the squared error in noise estimation. 

The ``sig-cov'' method of \cite{Yoshioka18} was used when computing the spatial covariance matrices. 
The gain adjustment technique of \cite{Yoshioka18} was also applied to reduce insertion errors.

\subsection{Details}
\label{subsec: train}
\vspace{-.4em}

We built an unmixing transducer by using a three-layer 1024-unit BLSTM. 
Input features were transformed by a 1024-unit projection layer with ReLU nonlinearity before being fed to the BLSTM. 
On top of the last BLSTM layer, there was a three-head fully connected sigmoid layer, where each head produced spectral masks for either speech or noise. 
Each of the heads consisted of 257 units, each uniquely associated with a particular frequency bin. 

567 hours of  speech mixture data were created for training. 
Source speech signals were taken from WSJ SI-284 and LibriSpeech. 
Each training sample was created as follows. 
First, the number of speakers (1 or 2) was randomly chosen. 
For the two-speaker case, the start and end times of each utterance was randomly determined so that we have 
a balanced mix of the four configurations described in \cite{Yoshioka18}. 
The source signals were reverberated with the image method~\cite{Allen79}, mixed together in the two-speaker case, and 
corrupted by additive noise. 
The multi-channel additive noise signals were simulated by assuming a spherical isotropic noise field. 
The generated training samples were clipped to 10 seconds. 
% Each WSJ utterance was used five times. 

Distributed training with 1-bit SGD~\cite{Seide14}  was performed on 16 GPUs by uisng Microsoft Cognitive Toolkit. 
The learning rate started from $2.0 \times 10^{-4}$  and divided by 10 after 150 epochs. 
The model was saved after each epoch. The model snapshot with the lowest validation loss was picked after convergence.

At test time, two additional tricks were utilized. 
First, for the two network heads producing the speech masks, 
we estimated the direction of signal arrivals (DOAs). 
When the DOA difference was less than 15 degrees, we assumed that there were actually only one speaker and thus
merged the masks while zeroing out the masks of the less significant head.
Secondly, for each time frequency bin, the three masks were normalized to sum to one.

%\subsubsection{Mask consolidation}

\section{Meeting Transcription Experiments}

\subsection{System build}
\vspace{-.4em}

We developed a meeting transcription system with a seven-channel circular array
by using the unmixing transduer described above. 
The system consists of three kinds of modules, each performing dereverberation, 
speech separation, or ASR. 
The dereverberation module estimates a multi-input multi-output dereverberation filter for converting a seven-channel microphone array signal to a less reverberant one with seven 
channels~\cite{Yoshioka12c}. 
The dereverberation filter was updated every second. 
%This attenuates the acoustic variations caused by reverberation and thus helps both speech separation and ASR. 
Then, the unmixing transducer transforms
the dereverberated seven-channel audio into two-channel separated speech streams. 
The model was built as per Section \ref{subsec: train}. 
Each output signal from the unmixing transducer was provided to an ASR back-end that performs segmentation and recognition.

For ASR, we trained an acoustic model on $\sim$7K hours of spontaneous speech audio, which were collected from various sources, both public (e.g., Switchboard and Fisher) and private (e.g., Microsoft Research lecture talks). 
The audio quality was not consistent due to 
the effects of noise, channel, and so on, which 
seemed to improve the robustness of the acoustic model.
The model input was 40-channel mel filterbank energies compressed with 10th-root nonlinearity.
The model consisted of four 1024-unit LSTM layers. 
It was trained with a cross entropy criterion, followed by sequence training. 
Decoding was performed with a dictionary of $\sim$240K words and our internal trigram language model built for conversational tasks.

\subsection{Task}
\vspace{-.4em}

To evaluate the meeting transcription system, six meetings were recorded at our speech group and professionally transcribed. 
The meetings took place in several different rooms and lasted for 30 minutes to an hour. 
%Many nonnative speakers were involved in our meetings. The number of attendees varied from three to seven. 
The recordings were made with both headset microphones and a seven-channel circular microphone array. 
They were manually segmented and transcribed. 
The transcribers were allowed to access both types of microphones. 
%It is noteworthy that, in some meetings, the microphone array was unintentionally put near a video conference system that was 
%always generating small noise. The SNRs for those meetings tended to be very low. 
The average overlap rate was 14.7\%, which was twice as high as that of AMI~\cite{Cetin06}. 
While the overlap rate value varies depending on the annotation policy, based on an informal inspection, we feel our meetings had noticeably more overlaps than those of AMI. 
The more frequent overlaps may be attributed to the fact that the participants in our meetings knew each other well and were discussing 
work-related topics. 

The system outputs were scored with asclite tool~\cite{Fiscus06}, which aligns multiple hypotheses against
multiple references. Tighter reference segmentations were used when calculating a word error rate (WER) for single-speaker segments so 
as not to discard segments that had overlaps only in nonspeech frames. Note that this might have resulted in a slight WER overestimation
because asclite makes use of the reference time stamps to find segments for alignment.

\subsection{Results}
\vspace{-.4em}

\begin{table}
\centering
\caption{\%WER of different front-ends.}
\label{tab: results}
\vspace{-1em}
\begin{tabular}{|l||c|c|}\hline
\multirow{2}{*}{System} & \multicolumn{2}{c|}{Overlapped segments} \\ 
 & Included & Excluded \\ \hline\hline
 No processing (mic0) & 44.6 & 40.9 \\
  Dereverb.~\cite{Yoshioka12c} & 42.1 & 38.7 \\
 ~~~+BeamformIt~\cite{Anguera07}  & 43.2 & 40.6\\
 ~~~+MaskBF~\cite{Boeddeker18}  & 37.9 & 32.8 \\ \hline
 ~~~+\textbf{Unmix. Trans. (proposed)} &  \textbf{33.8} & \textbf{30.4} \\
 ~~~+UT trained only on WSJMix &  34.2 & 30.8 \\ 
 ~~~+UT without noise channel &   36.8 & 34.5 \\ \hline
\end{tabular}
\vspace{-1.5em}
\end{table}

Table \ref{tab: results} lists the WERs obtained with different front-ends including our system. 
Without microphone array processing, the WER was 44.6\%. 
Dereverberation improved the recognition accuracy by 5.6\% relative. 
BeamformIt~\cite{Anguera07}, the default beamformer used by Kaldi's recipe for AMI~\cite{KaldiPaper}, provided no improvement.
A neural mask-based beamformer, which yielded state-of-the-art results in both CHiME-3/4~\cite{Heymann17} and more practical large vocabulary settings~\cite{Boeddeker18}, was also examined. 
Here, we used the best model we obtained in \cite{Boeddeker18} and applied it to our meeting data by using a 2.4-sec sliding window. 
This improved the performance by 10.0\%, achieving a WER of 37.9\%. 
However, this beamformer provided a 15.2\% gain for single-speaker segments, indicating that this method 
was not effective at handling overlaps. 
Nevertheless, we used this single-output beamformer as our baseline because 
no speech separation method was previously applied to meetings with unknown and varying numbers of attendees.

The proposed system achieved a WER of 33.8\%, outperforming the strong baseline system by 10.8\%. 
It is noteworthy that the gain was 7.3\% when the overlapped segments were excluded. 
This means that, while the proposed approach provided a modest improvement even for single-speaker segments, 
its advantage was prominent in overlapped segments. 
When the unmixing transducer was trained only on the WSJ-derived data, which amounted to 219 hours, 
the recognition accuracy slightly deteriorated. 
When the network estimated only speech masks, i.e., when the SSN architecture was not used, 
the recognition performance was degraded to 36.8\%.
The degradation was profound especially in single-speaker segments, indicating 
the importance of explicitly estimating the  noise masks. 
Overall, the proposed meeting transcription system, comprising the dereverberator, unmixing transducer, and ASR back-ends, 
improved the WER by 24.2\% compared with the single distant microphone system.

\section{Conclusion}
In this paper, we described a meeting transcription system that can handle speech overlaps. 
The system is based on the unmixing transducer, a novel signal processing module for 
converting multi-channel audio signals into a fixed number of separated speech streams. 
We implemented it by using a windowed BLSTM. 
The SSN architecture was proposed to effectively leverage beamforming capability.
Significant gains in meeting transcription performance were obtained compared with a strong neural mask-based beamformer. 
Further results on both public and private data will be reported in a follow-up paper. 

As far as we know, this is the first overlapped speech recognition system that has been demonstrated to 
work for actual unconstrained meetings.
We believe the proposed approach is promising and anticipate further investigation in this direction.

\newpage

%\section{Acknowledgements}

%The authors thank Zhuo Chen and Xiong Xiao for their feedback on this work. 

\bibliographystyle{IEEEtran}

\bibliography{my_references}

% \begin{thebibliography}{9}
% \bibitem[1]{Davis80-COP}
%   S.\ B.\ Davis and P.\ Mermelstein,
%   ``Comparison of parametric representation for monosyllabic word recognition in continuously spoken sentences,''
%   \textit{IEEE Transactions on Acoustics, Speech and Signal Processing}, vol.~28, no.~4, pp.~357--366, 1980.
% \bibitem[2]{Rabiner89-ATO}
%   L.\ R.\ Rabiner,
%   ``A tutorial on hidden Markov models and selected applications in speech recognition,''
%   \textit{Proceedings of the IEEE}, vol.~77, no.~2, pp.~257-286, 1989.
% \bibitem[3]{Hastie09-TEO}
%   T.\ Hastie, R.\ Tibshirani, and J.\ Friedman,
%   \textit{The Elements of Statistical Learning -- Data Mining, Inference, and Prediction}.
%   New York: Springer, 2009.
% \bibitem[4]{YourName17-XXX}
%   F.\ Lastname1, F.\ Lastname2, and F.\ Lastname3,
%   ``Title of your INTERSPEECH 2018 publication,''
%   in \textit{Interspeech 2018 -- 19\textsuperscript{th} Annual Conference of the International Speech Communication Association, September 2-6, Hyderabad, India Proceedings, Proceedings}, 2018, pp.~100--104.
% \end{thebibliography}

\end{document}